\documentclass[]{aa}
\usepackage{ulem}
\usepackage{amsmath}
\usepackage{graphicx}
\usepackage{natbib}
\usepackage{multicol}
\usepackage{longtable}
\usepackage{graphicx}
\usepackage{soul}
\usepackage{supertabular,booktabs}
\usepackage{sidecap}
\usepackage{enumitem}
\usepackage{txfonts}
\usepackage{epstopdf}
\usepackage{tabularx}
\usepackage{longtable}
\usepackage{ltablex}

\usepackage[pdftex]{xcolor}
\usepackage[]{natbib}
\def\Msun{$M_\odot$}

\begin{document}

\title{
Does the i-process operate at nearly solar metallicity?
}
%Evidence for the i-process operating at nearly solar metallicity 
%\subtitle{High-resolution re-analysis  
%of LAMOST barium stars}
%A word of caution about machine-learning studies using Sr lines

\author{D. Karinkuzhi
          \inst{1,2}
          \and
          S. Van Eck\inst{2}
          \and 
          S. Goriely\inst{2}
          \and
          L. Siess\inst{2}
           \and
          A. Jorissen\inst{2}
                    \and
          A. Choplin\inst{2}
          \and
          A. Escorza\inst{3}
          \and
          S. Shetye\inst{4}
          \and
          H. Van Winckel\inst{5}
          }

 \institute{Department of Physics, University of Calicut, Thenhipalam, Malappuram 673635, India 
           \and
            Institut d'Astronomie et d'Astrophysique, Universit\'e Libre de Bruxelles (ULB) C.P. 226, B-1050 Bruxelles, Belgium
              \and
      European Southern Observatory, Alonso de Córdova 3107, Vitacura, Casilla 19001, Santiago de Chile, Chile
      \and
      Institute of Physics, Laboratory of Astrophysics, École polytechnique fédérale de Lausanne (EPFL), Observatoire de Sauverny, 1290 Versoix, Switzerland
\and
Instituut voor Sterrenkunde, KULeuven, Celestijnenlaan 200D, 3001 Leuven, Belgium
             }

   \date{Received X, 2023; accepted Y, 2023}
   
\abstract
% context (optional)
{    
A sample of 895 s-process-rich candidates has been found among the 454\;180 giant stars surveyed by LAMOST at low spectral resolution ($R \sim 1800$).   
%A sample of 895 s-process-rich candidates out of 454\;180 giant stars surveyed by LAMOST at low spectral resolution ($R \sim 1800$) has been reported by Norfolk et al. (2019; N19). 
 In a previous study, taking advantage of the higher resolution ($R \sim 86\;000$) offered by the HERMES-Mercator spectrograph, we performed the 
re-analysis of 15 among the brightest stars of this sample.
Among these 15 programme stars, having close-to-solar metallicities,
11 showed mild to strong heavy element overabundances. 
The nucleosynthetic process(es) at the origin of  
these overabundances were however not questioned in our former study.
}
%Aim
{
We derive the abundances in s- and r-process elements of the
15 targets 
in order to investigate whether some stars also show an i-process signature, as sometimes found in their lower metallicity counterparts (namely, the Carbon-Enhanced Metal-Poor (CEMP)-rs stars).
}
    % methods
{Abundances are derived from the high-resolution HERMES spectra for  
Pr, Nd,
Sm, and
Eu,
using the  TURBOSPECTRUM radiative transfer  LTE code   with  MARCS  model atmospheres.  
}
%results
{Using the new classification scheme proposed in our recent study
we find that 
%some 
two stars show 
overabundances in both s- and r-process elements well above the level expected from the Galactic chemical evolution, an analogous situation to the one of CEMP-rs stars at lower metallicities. We compare the abundances of the 
%three 
most enriched stars with the nucleosynthetic predictions from the STAREVOL stellar evolutionary code and find abundances compatible with an i-process occurring in AGB stars.
%at higher metallicities.
    % results
}
%Conclusions
{Despite a 
%more sophisticated classification scheme using a 
larger number of heavy elements to characterize the enrichment pattern, the limit between CEMP-s and CEMP-rs stars remains fuzzy. 
It is however interesting to note that
an increasing number of extrinsic stars are found to 
have abundances better reproduced by an i-process pattern even at close-to-solar metallicities.
}
    % conclusion (optional)
    
\keywords{Nuclear reactions, nucleosynthesis, abundances -- Stars: AGB and post-AGB -- binaries: spectroscopic}

\maketitle

\section{Introduction}

The origin of the peculiar chemical composition of  CEMP-rs stars (carbon-enhanced-metal-poor stars enriched in {\it both} s and r elements)
is an unsolved problem. 
Some of the scenarios that have been invoked to explain their overabundances \citep{Jonsell2006,Masseron2010,Hampel2016} include (i) a primordial origin (the pollution of the birth cloud by an r-process source), (ii) a pollution of the binary by a third massive star (triple system), (iii) a pollution by the primary (type 1.5 supernova or accretion-induced collapse), or finally, (iv) the intermediate neutron-capture process or i-process. This nucleosynthesis process leads to 
 neutron densities of the order of $N_n \sim ~10^{15}$ cm$^{-3}$, 
intermediate between those of the s-process ($N_n \sim ~10^{8}$ cm$^{-3}$) 
and those required by the r-process ($N_n \gg ~10^{20}$ cm$^{-3}$).  Recently, \citet{karinkuzhi2021}  studied a sample of 
CEMP-s and rs stars in our Galaxy and suggested that the abundance pattern in CEMP-rs stars can be produced by a TP-AGB star experiencing i-process nucleosynthesis after proton ingestion during its first convective thermal pulse, and transferring this material to a close-by companion. From a  comparison of the abundance profiles of CEMP-s and CEMP-rs stars, \citet{karinkuzhi2021} noticed an abundance continuum rather than dichotomic patterns. Hence, the i-process could be a manifestation of the s-process at low ([Fe/H]$\sim -2$) metallicities, when proton injection takes place. 
Our previous identification and 
analysis \citep{Karinkuzhi2018}  of a star enriched in s and r elements at 
%comparatively 
higher metallicity ([Fe/H]~$\sim -0.7$) hints at an i-process which would not be limited to metal-poor environments. 
The literature contains other pieces of evidence in that direction, such as the barium star subsamples of  
 \cite{Cui-2014} and \cite{denHartogh2022},
and the born-again phase of the {\it Sakurai} object which also seems to display an i-process pattern \citep{Herwig2011}. These observations show that it seems to be possible for a given object 
to produce this rs pattern without
resorting to a "two-event" scenario (independent and cumulative pollutions by an r-process and an s-process).

In this paper we use sample of 
15 stars from \citet{Norfolk2019}
reanalyzed in
\citet[][hereafter K21]{Karinkuzhi2021b} 
to check for potential hybrid profiles at close-to-solar metallicities.
This paper is organized as follows. Section~\ref{Sect:sample} describes the sample selection and Sect.~\ref{Sect:parameters-abundances} the method used to derive the atmospheric parameters and abundances. 
In Section~\ref{Sect:classification} 
 we classify the stars 
on the basis of their [s/r] ratio, while in Sect.~\ref{Sect:7-elements} we use a classification based on a larger number of heavy elements.
Section~\ref{Sect:nucleosynthesis} presents AGB models and nucleosynthesis reproducing the measured abundances with an i-process, triggered by a  proton-injection episode (PIE) very similar to the one found in metal-poor stellar models.
In Section~\ref{Sect:sandr} we show that the i-process provides a better agreement with the measered abundances than a superposition of two independent s- and r-pollutions.
Conclusions are presented in Sect.~\ref{Sect:conclusions}.

\section{Sample selection}
\label{Sect:sample}
Our %initial 
sample 
%\citep{karinkuzhi2021b} 
consists in 
15 bright barium stars selected from the 895 objects from \citet{Norfolk2019} and for which high-resolution HERMES \citep{Raskin2011} spectra could be obtained. A first analysis of this sample was presented in K21.
The metallicities 
%of the sample 
range from [Fe/H]~$=0.02$ to $-0.61$. 
Among the 15 programme stars, 4 show no s-process overabundances ([X/Fe] $<$ 0.2 dex), 8 show mild s-process overabundances (at least three heavy elements with $0.2 \le$ [X/Fe] $<$ 0.8), and 3 have strong overabundances (at least three heavy elements with [X/Fe] $\ge$ 0.8), as listed in Table.~\ref{Tab:programme_stars}.

The binarity of these stars was investigated
by K21 from which it appears that 2 out of the 3 strong barium stars show a clear binary signature from radial-velocity (RV) variations. The results for the other classes are intriguing, however, since only 1 out of the 8 mild barium stars diagnosed exhibit statistically significant RV variations, and on the opposite, 2 out of the 4 ‘no-s’ stars show a binary signature.

\section{Derivation of atmospheric parameters and abundances}
\label{Sect:parameters-abundances}

The atmospheric parameter derivation
is presented in K21. The stellar parameters are repeated here 
%in the first columns of
(Table~\ref{Tab:programme_stars}).

 \begin{table*}
\caption{Programme stars, adopted atmospheric parameters from K21 and abundance profile distances to the r-process (Sect.~\ref{Sect:7-elements}). $\xi$ is the microturbulence velocity. 
For the criteria used to classify stars as 'no s-process', 'mild s-process enrichment' and 'strong s-process enrichment', see Sect.~\ref{Sect:classification} of \citet{Karinkuzhi2021b}. 
In the column 'Remark', we identify two stars where a Li abundance could be derived ('Li' if the Li abundance is subsolar and 'Li-rich' if it is super-solar).
}
\label{Tab:programme_stars}
\begin{tabular}{lllrccccccc}
\hline
\\
Name &  \multicolumn{1}{c}{$T_{\rm eff}$}&\multicolumn{1}{c}{$\log g$}  & \multicolumn{1}{c}{[Fe/H]}& $\xi$ &Signed & RMS & \multicolumn{1}{c}{Remark}&  \\
     &\multicolumn{1}{c}{(K)}       &\multicolumn{1}{c}{(cm s$^{-2}$)}  &\multicolumn{1}{c}{(dex)}  &  (km s$^{-1}$) & distance & distance &\multicolumn{1}{c}{}  &\\
\hline\\
\multicolumn{8}{c}{\bf no s-process enrichment}\\
HD 7863&4637 $\pm$ 64 & 2.29 $\pm$ 0.40& $-$0.07 $\pm$ 0.05&1.26 $\pm$ 0.10              & 0.37  & 0.43&  && \\
HIP 69788&5127 $\pm$ 11& 3.90 $\pm$ 0.14& $-$0.04 $\pm$ 0.04&0.61 $\pm$ 0.10             & 0.65 & 0.70 &  &&  \\
TYC 3144$-$1906$-$1&4136 $\pm$ 64 & 1.89 $\pm$ 0.50& $-$0.13 $\pm$ 0.10 & 1.37 $\pm$ 0.04& 0.41 & 0.45 & Li&&  \\
TYC 4684$-$2242$-$1&4651 $\pm$ 20 & 2.70 $\pm$ 0.14 & $-$0.05 $\pm$ 0.07& 1.15 $\pm$ 0.05& 0.28 & 0.36 &  && 
\\
\multicolumn{8}{c}{\bf mild s-process enrichment}\\
BD $-07^\circ$ 402 &4654 $\pm$ 6 & 2.62 $\pm$ 0.19& $-$0.11 $\pm$ 0.05&1.22 $\pm$ 0.10 & 0.46 & 0.51 & Li-rich&& \\
BD $+44^\circ$ 575 & 4175 $\pm$ 6 & 1.50 $\pm$ 0.19& $-$0.45 $\pm$ 0.05& 1.60 $\pm$ 0.10 & 0.43& 0.47 & &&\\
TYC 22$-$155$-$1&4704 $\pm$ 9& 3.10 $\pm$ 0.32 &$-$0.20 $\pm$ 0.10 & 1.04 $\pm$ 0.05& 0.20 & 0.32 & && \\
TYC 2913$-$1375$-$1&4757 $\pm$ 69& 2.00 $\pm$ 0.30 & $-$0.61 $\pm$ 0.11 & 1.45 $\pm$ 0.05& 0.24& 0.32 & && \\
TYC 3305$-$571$-$1&4816 $\pm$ 3 & 2.76 $\pm$ 0.16 & $-$0.05 $\pm$ 0.08 & 1.31 $\pm$ 0.04& 0.48& 0.53 & && \\
TYC 752$-$1944$-$1&5069 $\pm$ 25   & 2.94 $\pm$ 0.05 & $-$0.08 $\pm$ 0.08 &1.33 $\pm$ 0.04& 0.70 & 0.74 & &&\\
TYC 4837$-$925$-$1&4679 $\pm$ 34   & 2.16 $\pm$ 0.29 & $-$0.27 $\pm$ 0.07& 1.30 $\pm$ 0.04& 0.37 & 0.43 & &&\\
TYC 3423$-$696$-$1&5042 $\pm$ 64 & 3.66 $\pm$ 0.30 & 0.02 $\pm$ 0.08 & 0.96 $\pm$ 0.04& 0.35& 0.41 &  &&  \\
\multicolumn{8}{c}{\bf strong s-process enrichment}\\
TYC 2250$-$1047$-$1&5335 $\pm$ 25& 3.71 $\pm$ 0.18 & $-$0.55 $\pm$ 0.12 & 1.45 $\pm$ 0.05& 0.56 & 0.64 & &&  \\
TYC 2955$-$408$-$1&4716 $\pm$ 64 & 2.49 $\pm$ 0.3 & $-$0.39 $\pm$ 0.08 & 1.25 $\pm$ 0.04& 0.74 & 0.79 & &&   \\
TYC 591$-$1090$-$1&5267 $\pm$ 36   & 3.68 $\pm$ 0.50 & $-$0.30 $\pm$ 0.12& 1.18 $\pm$ 0.06& 0.63&
0.69 & && \\
\hline
\end{tabular}
\end{table*}

\setlength{\tabcolsep}{6pt}

\begin{figure}
\includegraphics[width=9.0cm]{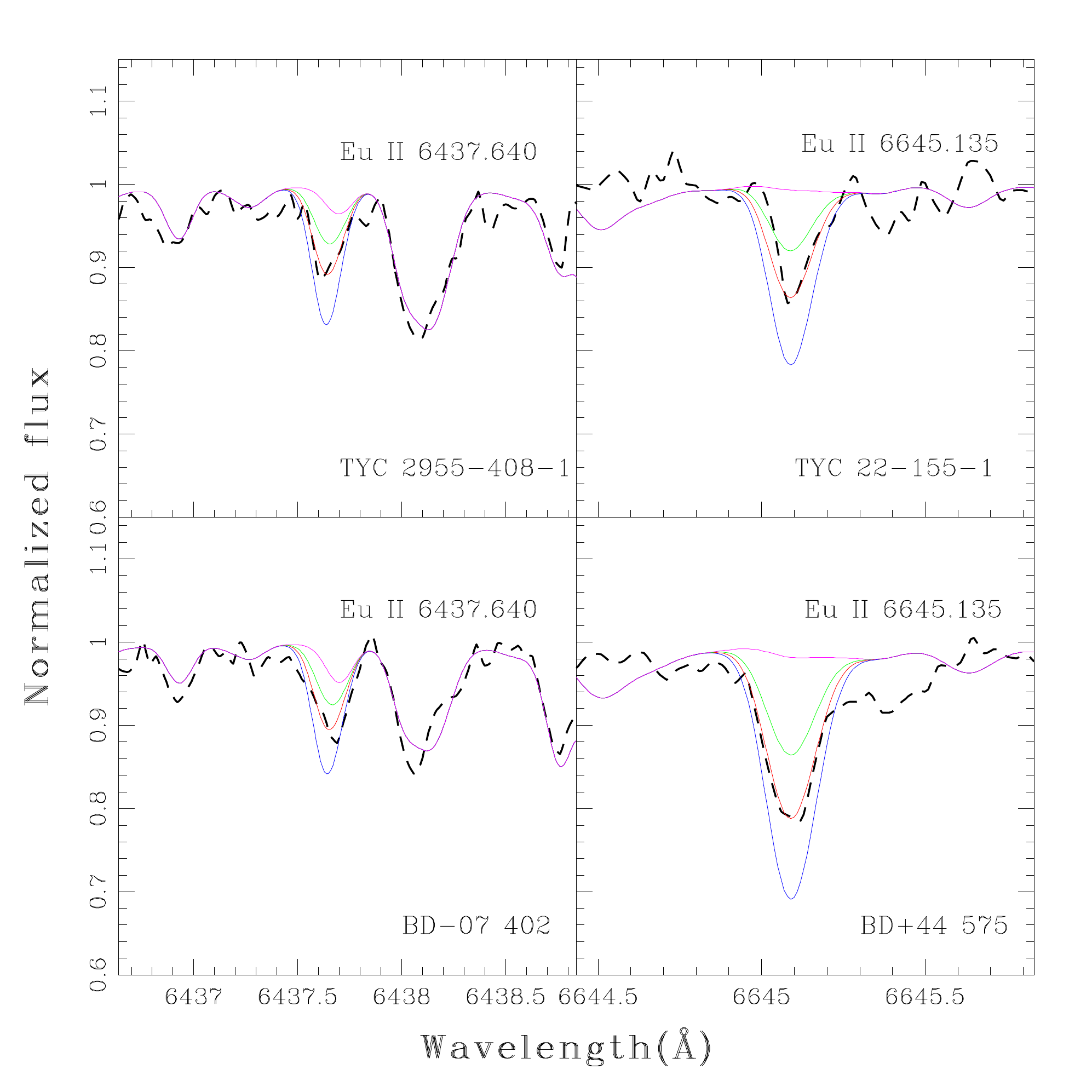}
\caption{Spectral fitting of the
6437.640 and 6645.135 \AA~
\ion{Eu}{II} lines for four sample stars. Red lines correspond to spectral syntheses with the abundance for Eu listed in Table~\ref{Tab:abundances}. Blue and green lines represent syntheses with abundances deviating by $\pm$0.3~dex from the adopted abundance. The magenta line corresponds to the synthesis with a null abundance for Eu.
\label{Fig:Eu} }
\end{figure}

Abundances were derived through spectral synthesis using the
LTE TURBOSPECTRUM code \citep{Alvarez1998}.
We used solar abundances from \citet{asplund2009} and the line lists 
of \citet{heiter2015,heiter2020} 
as in \cite{Karinkuzhi2018,karinkuzhi2021}. 
Table~\ref{Tab:abundances} presents the derived Pr, Nd, Sm, and Eu abundances, while we refer to Tables A.1 and A.2 of K21
for the remaining abundances. The complete abundance profiles are also presented in Fig.~\ref{Fig:pattern2} and \ref{Fig:pattern3}.
\citet{mashonkina-2000} and \citet{mashonkina2014} studied the NLTE corrections for the \ion{Eu}{II} 6645.135 \AA\ line and showed that they are negligible ($\approx$ 0.06 to 0.08 dex for metallicities higher than $-3.0$).
For the other elements, we could not find information about NLTE corrections at the metallicities of our objects.
The isotopic shifts and hyperfine splitting (HFS) of the atomic lines have been included while deriving the Eu abundances. For other elements the isotopic shifts and HFS splitting are not available for the lines we used. 
Fig.~\ref{Fig:Eu} presents the spectral fitting of Eu lines in a few sample stars.
Though many Nd lines are available throughout the spectral coverage, we used the \ion{Nd}{II} lines in the range 5200 -- 5400~\AA\, since they are relatively free from molecular blends. For Pr and Sm, we used all the measurable lines as listed in \citet{Karinkuzhi2018,karinkuzhi2021}. The details about the error estimates are presented in K21. 
The resulting changes in the abundances by varying the atmospheric parameters $T_{\rm eff}$, $\log g$, microturbulence $\xi$, and [Fe/H] by 100~K, 0.5, 0.5~km/s and 0.5 dex, respectively are presented in Table~\ref{Tab:uncertainties}. The final errors are calculated using Eqs.~1 and 2  of K21 and are presented in Table~\ref{Tab:abundances}. 

\begin{table}
\caption{Sensitivity of the abundances ($\Delta \log \epsilon_{X}$) with variations of the atmospheric parameters (considering the atmospheric parameters of BD $-07^\circ$ 402 ). }
\label{Tab:uncertainties}
\begin{tabular}{crrrr}
\hline
\\
       & \multicolumn{4}{c}{$\Delta \log \epsilon_{X}$} \\
        \cline{2-5}
        \\
 Element&   $\Delta T_{\rm eff}$ & $\Delta \log g$  &$\Delta$ [Fe/H]& $\Delta \xi_t$   \\
&   ($+$100 K) & ($+$0.5) & ($+$0.5  & ($+$0.5 \\
   &        &         & dex)& km~s$^{-1}$) \\
\hline\\

Nd & 0.00 & 0.13 &  0.07 &  0.00 \\
Pr & 0.03 & 0.19 &  0.22 & -0.01 \\
Sm & 0.06 & 0.30 & 0.15  & 0.02 \\
Eu & 0.05 & 0.22 &  0.03 &  0.06 \\

\hline
\end{tabular}
\end{table}

\section{Classification based on [Ba/Eu] and [La/Eu] abundance ratios}
\label{Sect:classification}

\begin{figure}[t!]
\includegraphics[width=10cm]{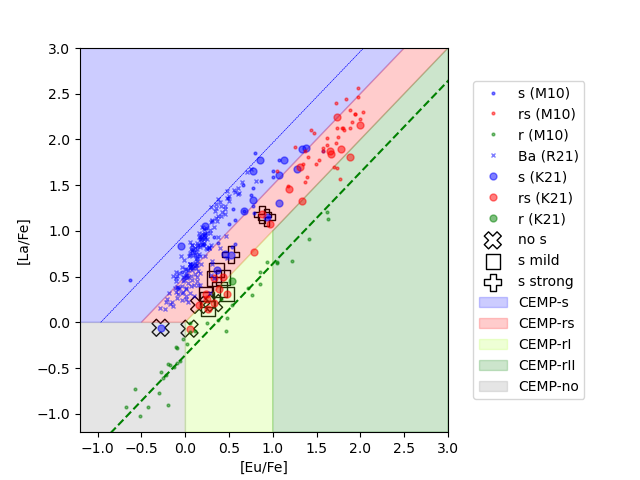}
\caption{[La/Fe] as a function of [Eu/Fe]. CEMP-s stars, CEMP-rs stars, and r-process enriched stars from  \citet{Masseron2010} and \citet{Karinkuzhi2018, karinkuzhi2021} as well as the Ba stars  of \citet{roriz2021} are included in the figure. 
The dashed green line corresponds to abundance-ratio scaling with a pure solar r-process \citep{Goriely1999}, whereas the continuous blue line corresponds to s-process nucleosynthesis abundance ratio scaling with the predictions from the 5th pulse of a 1.5 M$_\odot$ star at [Fe/H] = -1.  }
\label{Fig:2}
\end{figure}

\begin{figure}[t!]
\includegraphics[width=10cm]{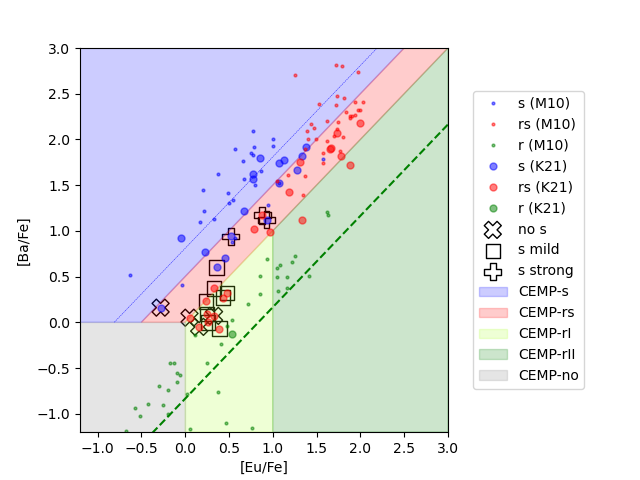}
\caption{
Same as Fig.~\ref{Fig:2}, but with [Ba/Fe] as a function of [Eu/Fe].}
\label{Fig:1}
\end{figure}

Various abundance thresholds have been used for the  classification of CEMP-s and -rs stars 
%we refer the reader to the 
 \citep[see discussion in Sect.~5 of][]{karinkuzhi2021}. 
In Fig.~\ref{Fig:2} (resp., Fig.~\ref{Fig:1}), typical (s,r) planes are presented, using lanthanum (resp., barium) as a prototype s-element, and europium as a representative r-element. The different colored regions correspond to commonly accepted limits 
\citep{beer2005} for CEMP-s stars ([Ba/Eu] or [La/Eu] > 0.5, blue region), for
CEMP-rI stars ([Ba/Eu] or [La/Eu] < 0, 0 < [Eu/Fe] < +1, light green region), and for CEMP-rII stars ([Ba/Eu] or [La/Eu] < 0, [Eu/Fe] > +1, dark green region). 
Finally, the intermediate region between the blue CEMP-s and the green CEMP-r regions has been colored in red.  
A growing number of stars enriched in both r and s elements (CEMP-rs) has been found in the past decade to occupy this region. 
In Figs.~\ref{Fig:2} and ~\ref{Fig:1}, CEMP-r, -s and -sr stars from \cite{Masseron2010}, \citet{Karinkuzhi2018}, \citet{karinkuzhi2021} are shown as green, blue and red dots, respectively. In addition, barium stars analyzed by \citet{roriz2021} (for which the barium abundance is not available) are plotted as small blue crosses in Fig.~\ref{Fig:2}.

The majority of barium stars analysed by 
\citet{roriz2021} and of CEMP-s stars
is located as expected in the blue-shaded region. However, we note that especially at low enrichment levels ([La/Fe] and [Eu/Fe] $<0.7$), many barium and CEMP-s stars occupy the red-shaded region. 
Actually if the enrichment is low, the pristine composition might not be erased by the pollution from the companion and 
the current abundance profile  might also reflect an enrichment due to the Galactic chemical evolution.
Alternatively, it can be due to the difficulty to determine weak line abundances in blended spectra\footnote{We do not consider stars with low  enrichment levels ([X/Fe]$<0.2$ dex) as barium stars, given the typical  uncertainties affecting the abundance determinations.}.

The 15 stars of the current sample are identified in Fig.~\ref{Fig:2} with black symbols surrounding the corresponding dots: black crosses for the "no s-process enrichment", black squares for the "mild enrichment" and pluses for the "strong enrichment" of Table~\ref{Tab:programme_stars}. Interestingly, they all fall 
%close to the location of barium stars (blue region in Fig.~\ref{Fig:2}), but 
in the  CEMP-rs (red) or r-I (green) regions.
This is reminiscent of the barium stars analyzed by \citet{denHartogh2022}: these authors detect a 
signature of an i-process activation in about 25\% of the evolved companions of the barium star.

\section{Classification based on s- and r- elements}
\label{Sect:7-elements}
\begin{figure}
\includegraphics[clip,trim=1cm 0cm 0cm 0cm,width=10cm]{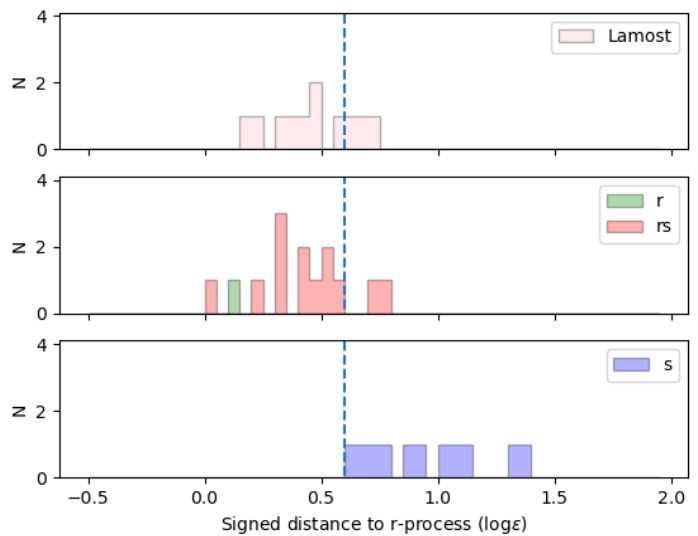}
\caption{{\it Upper panel}: Histogram of the signed distance (Eq.~\ref{Eq:dist-signed}) of the star sample of the present study  (considering only those with mild and strong enrichments, so 11 stars out of 15).
{\it Middle panel}: Same histogram for the \citet{Karinkuzhi2018} sample of CEMP-rs (red) and -r (green) stars.
{\it Lower panel}: Same histogram for the \citet{Karinkuzhi2018} sample of CEMP-s stars.
The dashed line represents the $d_{\rm S}=0.6 $ limit between CEMP-s and CEMP-rs stars, as defined in \citet{karinkuzhi2021}.}.
\label{Fig:signed-distance-histo}
\end{figure}

 We have used 7 elemental abundances
 (Y, Zr, Ba, La, Ce, Nd, Sm)
 to compute a signed distance to a reference abundance pattern, chosen as the solar r-process, scaled to the europium abundance of the star (see Eq.~3 of \citealt{karinkuzhi2021}):
 \begin{equation}
d_{\rm S} = \frac{1}{N}\sum_{x_i} (\log_{10} \epsilon_{x_i,\ast} - \log_{10} \epsilon_{x_i,\rm{norm(r,\ast)}})  
\label{Eq:dist-signed}
\end{equation}
where $\{x_1... x_N\}$ are the $N$ considered heavy elements, and
we use the usual notation $\log_{10} \epsilon_{x_i} = \log_{10} (n_{x_i}/n_{\rm H}) + 12$, with $n_{x_i}$ the number density of element $x_i$. We denote
 $\log_{10} \epsilon_{x_i,\ast}$  the abundance of element $x_i$ for the program stars,
and
$\log_{10} \epsilon_{x_i,\rm{norm(r,\ast)}}$  the standard r-process abundance profile $\log \epsilon_{x_i, r}$  normalized to the star abundance profile with respect to europium:
\begin{equation}
\log \epsilon_{x_i,\rm{norm(r,\ast)}} = \log \epsilon_{x_i, r} + (\log\epsilon_{Eu,\ast} - \log\epsilon_{Eu,r}),
\end{equation}
where the adopted r-process abundances $\log \epsilon_{x_i,r}$ are listed in Table~B.4 of \citet{karinkuzhi2021}.
 Figure~\ref{Fig:signed-distance-histo} presents the histogram of this signed distance, together with the limit adopted in \citet{karinkuzhi2021} to separate CEMP-s from CEMP-rs stars (at $d_S=0.6)$. This limit is certainly somewhat arbitrary. However, Fig.~\ref{Fig:signed-distance-histo} shows that most of our sample stars strongly enriched in heavy elements appear to have an rs pattern.

\section{Comparison with nucleosynthesis predictions}
\label{Sect:nucleosynthesis}
AGB nucleosynthesis predictions have been computed coupling
the STAREVOL code \citep{Siess2008} with an
extended reaction network of 1160 species linked by 2123 nuclear reactions.  
More details on the nuclear network and input physics can be found in \citet{Choplin21} and \citet{Goriely21}.
The solar abundances are taken from \citet{asplund2009} which correspond to a metallicity $Z = 0.0134$. 
The \citet{Reimers1975} mass loss rate with $\eta_R=0.4$ is used from the main sequence up to the end of core helium burning and the \citet{Vassiliadis1993} during the AGB phase.
Dedicated models with an initial mass of 2~\Msun\ and a metallicity of [Fe/H]=$-0.5$, close to the one derived from the observations, have been computed as explained below.

In the present calculations, a diffusion equation is used to compute the partial mixing of protons in the C-rich layers at the time of the third dredge-up (TDU). We follow  Eq.~(9) of \citet{Goriely18c} and use in our standard case the same diffusive mixing parameters in our simulations as in \citet{Shetye19}, {\it i.e.}, $f_{\rm env}=0.14$, $D_{\rm min} = 10^7\,{\rm cm^2\, s^{-1}}$ and $p = 1/2$, where $f_{\rm env}$ controls the extent of the mixing, $D_{\rm min}$ the value of the diffusion coefficient at the 
base of the envelope, and $p$ is a free parameter describing the shape of the diffusion profile. In addition to this diffusive mixing at the base of the envelope, a similar prescription is applied at the top of the thermal pulse where a coefficient $f_{\rm pulse}$ governs the diffusive transport. Two different values of the $f_{\rm pulse}$ coefficient, triggering an s-process or i-process nucleosynthesis,  are considered here, as explained below.

\begin{figure}
\includegraphics[clip,trim=2cm 1cm 0cm 0cm, width=10.0cm]{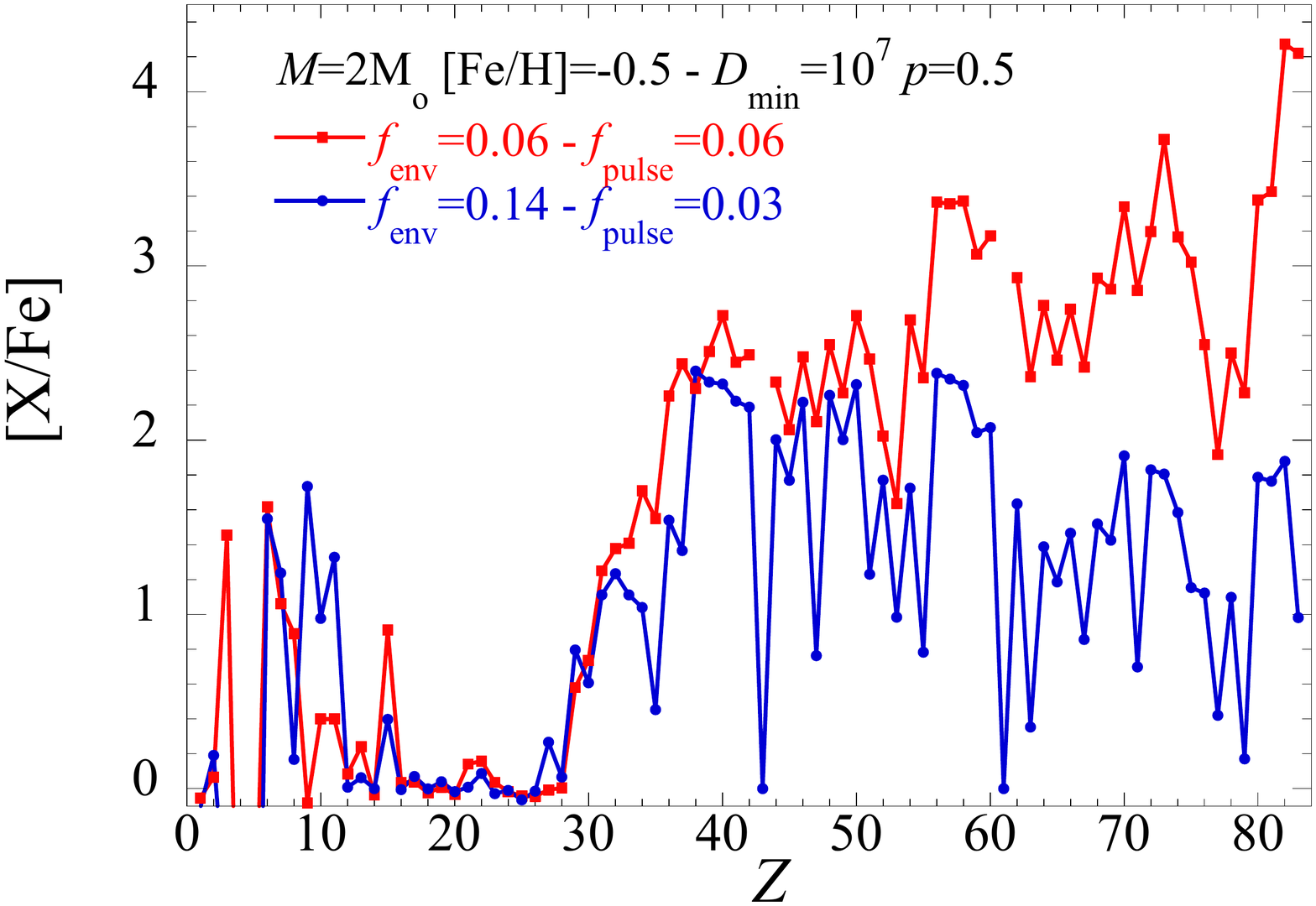}
\caption{Comparison of the final elemental surface distributions [X/Fe] (solid lines) obtained after the development of the neutron-capture processes in a 2~\Msun\ AGB star of metallicity ${\rm [Fe/H]}=-0.5$. The blue line is obtained with a diffusive mixing characterized by $f_{\rm env}=0.14$ and $f_{\rm env}=0.03$ and give rise to an s-process nucleosynthesis while the red line results from a mixing with $f_{\rm env}=0.06$ and $f_{\rm pulse}=0.06$ leading to a PIE, i.e., to an i-process nucleosynthesis.  
}
\label{Fig:i_vs_s} 
\end{figure}

When considering a relatively weak diffusive mixing at the top of the thermal pulse characterized by $f_{\rm pulse}=0.03$, the AGB phase of our 2~\Msun\ [Fe/H]=$-0.5$ model star is found to follow a standard evolution with the occurrence of 8 thermal pulses and a regular surface enrichment at the time of the TDU. In addition, the diffusive mixing at the bottom of the envelope gives rise to a radiative s-process during the interpulse phases similar to what is found in \citet{Goriely18c}. The corresponding abundance pattern found at the surface at the end of the AGB evolution is illustrated in Fig.~\ref{Fig:i_vs_s} (blue curve).

The ingestion of protons in the convective helium-burning zone can lead  to a rich i-process nucleosynthesis that can explain quite successfully the surface enrichment of CEMP-rs stars \citep[e.g.][]{Hampel2019,karinkuzhi2021,Choplin21}.
However, as detailed in \citet{Choplin22}, when no extra mixing is included, the proton ingestion event (PIE) is restricted to model stars of metallicity lower than typically ${\rm [Fe/H]} \sim -2$. When considering diffusive mixing below the envelope and at the top of the thermal pulse, it is found that a possible PIE can be triggered in a way very similar to those found in very metal-poor stars. In particular, mixing parameters $f_{\rm env}=0.06$ and $f_{\rm pulse}=0.06$ lead to a PIE at the time of the third thermal pulse, 
with a maximum neutron density reaching $10^{14} {\rm cm}^{-3}$. The resulting surface abundance distribution is shown in Fig.~\ref{Fig:i_vs_s} (red profile) and is rather similar to the one obtained in low-metallicity stars \citep{Choplin22}. 

\begin{figure}
\includegraphics[clip,trim=2cm 1cm 0cm 0cm,width=10cm]{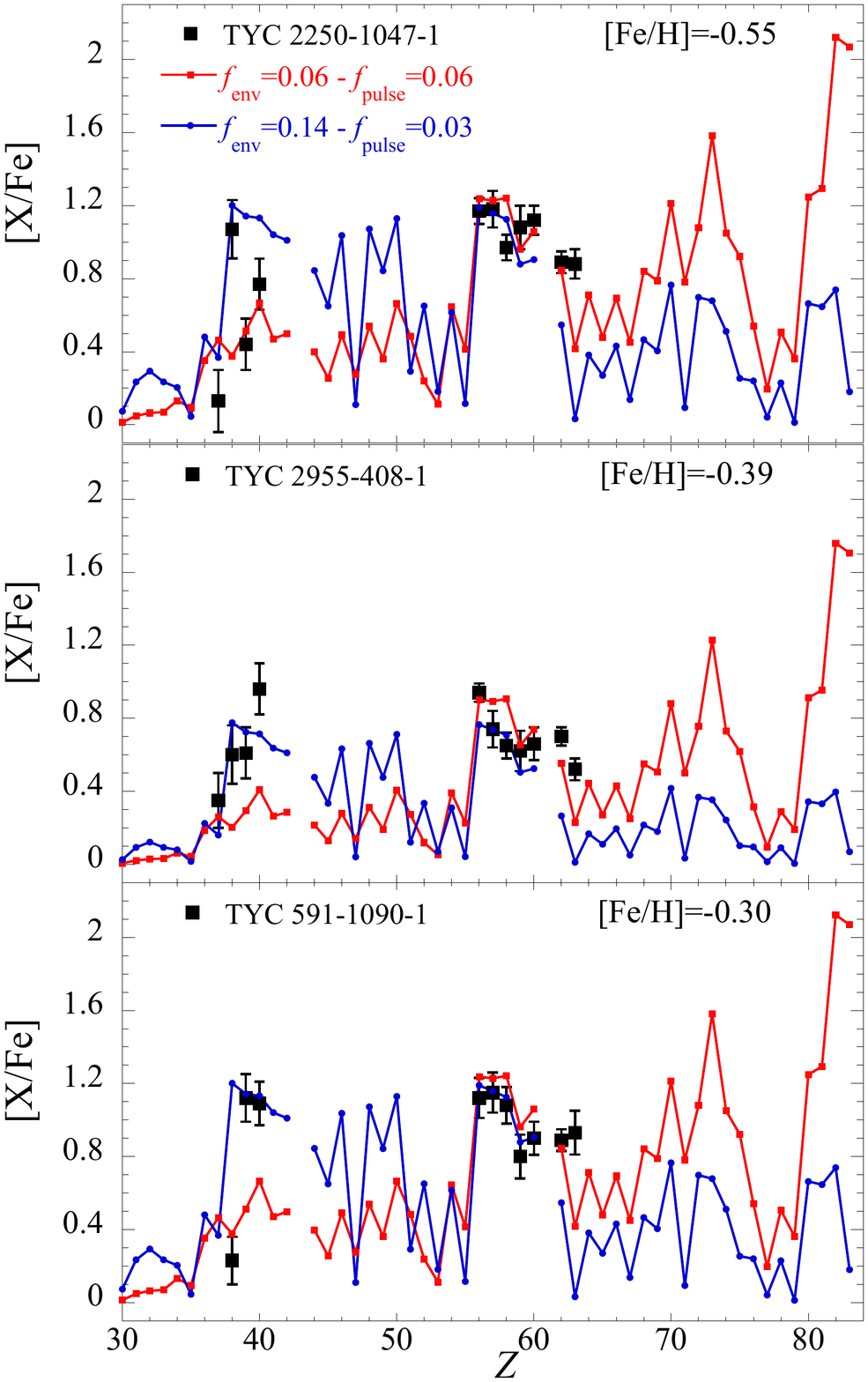}
\caption{
Comparison of the measured abundances with nucleosynthesis predictions from the STAREVOL code. 
The abundances obtained at the surface of the 2~\Msun\ [Fe/H]=$-0.5$ model with two different sets of the diffusive mixing parameters,
(the blue line corresponding to an s-process, and the red line to an i-process)
as shown in Fig.~\ref{Fig:i_vs_s},  are compared with the abundances of the 3 most enriched barium stars (black squares). 
}
\label{Fig:pattern}
\end{figure}

In summary, depending on the strength of the overshoot mixing imposed 
both at the bottom of the envelope and at the top of the pulse, either a 
standard s-process or an i-process can be simulated in a 2~\Msun\ 
[Fe/H]=$-0.5$ model star. Despite progress in multi-dimensional 
simulations of convective boundary mixing \citep[e.g., in low-mass stars,][]{Mocak2011, Herwig2014, Woodward2015}, no firm conclusion can be drawn on the strength of the mixing. Our 
approach is to explore these possibilities and compare our results with 
the present observations.

A particular attention is paid to the most strongly enriched s-process stars of Table~\ref{Tab:programme_stars},
because their abundances must reflect a genuine pollution and not the chemical enrichment of the Galaxy, contrarily to the situation that might prevail for stars with low-enrichments. 
In particular, TYC 2250-1047-1 and TYC 591-1090-1 have [Eu/Fe] of 0.88 and 0.93 dex, respectively, which cannot be explained by galactic chemical evolution alone. 
Virtually no star is found with [Eu/Fe]$>0.4$ among disk stars with [Fe/H]$>-0.4$
\citep{Tautvaisine-2021, VanderSwaelmen-2023}.
Therefore the abundance pattern of these two stars must result from extra pollution(s). TYC	2955-408-1, with [Eu/Fe]=0.52, is difficult to classify and the abundance of additional elements is needed to decide whether it is an i-process star or a barium star with a europium abundance corresponding to the high-europium tail of the Galactic distribution.
A comparison of their measured and predicted abundances is given in Fig.~\ref{Fig:pattern} where a dilution factor is applied to the surface abundances shown in Fig.~\ref{Fig:i_vs_s} to match the overabundances observed in the La-Ba region.  This dilution factor is required to simulate the mixing of the nucleosynthesis yields from the primary star onto its companion (assumed to be of the same initial composition). 

The overall accuracy of the model predictions can be quantified for each star through the reduced $\chi^2$ indicator:
%, defined as in \citet{Hampel2016}:
\begin{equation}
\chi^2= \frac{1}{N} \sum_X \frac{\left({[\rm X/Fe]_{obs}}-{[\rm X/Fe]_{mod}}\right)^2}{\sigma^2_{\rm X,obs}}\quad ,
\label{eq:chi2}
\end{equation}
where ${[\rm X/Fe]_{obs}}$ and ${[\rm X/Fe]_{mod}}$ are 
respectively the measured and predicted abundances of a given element $X$ and $\sigma_{\rm X,obs}$ is the associated uncertainty on the measured abundance. We consider the $N=10$ elements available in all stars, namely Sr,Y, Zr, Ba, La, Ce, Pr, Nd, Sm, and Eu. 

%\section{Reduced $\chi^2$ computation}

We report a typo in Eq.~6  of \citet{karinkuzhi2021}, where the normalizing factor $1/N$ was erroneously omitted, though it was included when computing the $\chi^2$ listed in the Table~1 of that paper \citep[see Corrigendum in][]{Karinkuzhi-corr-2023}. Equation~\ref{eq:chi2} above is the correct expression used to provide the numbers given in the tables of both papers.
\begin{table}
\caption{Reduced $\chi^2$ computed either (i) on the same 7  elements (Y, Zr, Ba, La, Ce, Nd, Sm) as for the signed or RMS distances  (where Eu is used as a normalizing element), (ii) based on the previous elements plus Eu and (iii) based on the 10 elements (Sr,Y, Zr, Ba, La, Ce, Pr, Nd, Sm, and Eu) with derived abundances in the present paper. 
For convenience, the signed distance (Eq.~\ref{Eq:dist-signed}) and the RMS distance 
\citep[Eq.~4 of][]{karinkuzhi2021}
are repeated in the last two columns. 
} 

\label{Tab:chi2}
\begin{tabular}{lcc c cc}
\hline\\
Name, &&&&&\\
Number        & \multicolumn{2}{c}{Reduced $\chi^2$}             &&     \multicolumn{2}{c}{Distance}   \\
\cline{2-3} \cline{5-6} 
of            &    s-pro &  i-pro &&     $d_S$ &  $d_{RMS}$ \\
elements            &  & && &    \\
\hline\\
\multicolumn{6}{l}{TYC 2250-1047-1} \\
7   &     10.9     &          2.6   && 0.56  &   0.64  \\
8   &     23.7  &             6.5   &&       &           \\
10  &     19.3  &             7.1   &&       &         \\
\\
\multicolumn{6}{l}{TYC 591-1090-1} \\
7  &    4.8    &            6.0   && 0.63  &   0.69 \\
8  &    11.2   &            7.6   &&       &        \\
10 &    14.6   &            6.4   &&       &        \\
\hline
\end{tabular}
\end{table}

In \cite{karinkuzhi2021}, reduced $\chi^2$ values for CEMP-s stars (when compared to an s-process) ranged between 2.7 and 10.9. Similarly, reduced $\chi^2$ values for CEMP-rs stars (when compared to an i-process) ranged between 1.3 and 10.6.
Here, for the 
%three 
two most-enriched stars 
(TYC 2250-1047-1 
%TYC 2955-408-1 
and TYC 591-1090-1),
we find reduced $\chi^2$ between 6.4 and 
7.6
%8.8 
(depending on the exact number (8 or 10) of considered chemical elements, but including Eu, see Table~\ref{Tab:chi2}) when comparing to an i-process, but in the range between 11.2 and 23.7 when comparing to an s-process.
Hence the chemical pattern of these 
%three 
two
most-enriched stars is better reproduced by an i-process than by an s-process.

We note that the first peak s-process element abundances (Sr, Y, Zr) are always  difficult to reproduce (either by an s- or i-process, Fig.~\ref{Fig:pattern}), and this strongly impacts the $\chi^2$.
The results from the $\chi^2$ are thus less clear with the 7 elements than with a metric using 8 (i.e. including Eu) or 10 elements (including Sr, Pr and Eu).
Considering Eu in addition to Sm is thus crucial when assessing an i- or r- contribution.
We note that among these two
stars, only TYC 2250-1047-1 
is classified as rs-enriched according to the signed distance criterion ($d_S < 0.6$). 
The reason partially lies in the fact that the 0.6 threshold on the signed distance is a somewhat arbitrary value and, as seen in Fig.~\ref{Fig:signed-distance-histo}, values up to $d_S=0.75$ might still be encountered for some CEMP-rs stars falling in the intermediate (red) regions of Fig.~\ref{Fig:2} and ~\ref{Fig:1}.

In summary, in this paper we use two different indicators to assess the s- or i-character of a measured stellar abundance profile: the distance indicator (using 7 heavy elements plus Eu which serves to normalize the measured Eu abundance to the solar-scaled "universal" r-process abundance; it has the advantage to be model-independent)
and the reduced $\chi^2$ (computed between the measured profile and a  nucleosynthetic predictions, either s or i). Table~\ref{Tab:chi2} shows that these two indicators are consistent when the comparison is based on the same set of chemical elements, 
as can be seen by comparing the reduced $\chi^2$ for 7 elements (6.0, 2.6) and the $d_S$ distance which uses the same 7 elements (0.63, 0.56) for the two stars (TYC 591-1090-1, TYC 2250-1047-1), ordered by increasing contribution of an i-process. 
The 8 and 10-element $\chi^2$ show that including Sr as well as Pr and Eu is very important to get clearly different s process $\chi^2$ and i-process $\chi^2$, and thus to get a clear diagnostic  concerning the s- or rs-character of the abundance profile.

We 
%also 
demonstrate in the next section that 
replacing the i-process pattern with
any superposition of two independent s- and r-pollutions deteriorates the agreement with the measured abundances.

%******

 \section{Testing the two independent pollutions scenario}
\label{Sect:sandr}

\begin{figure}
\includegraphics[width=9cm]{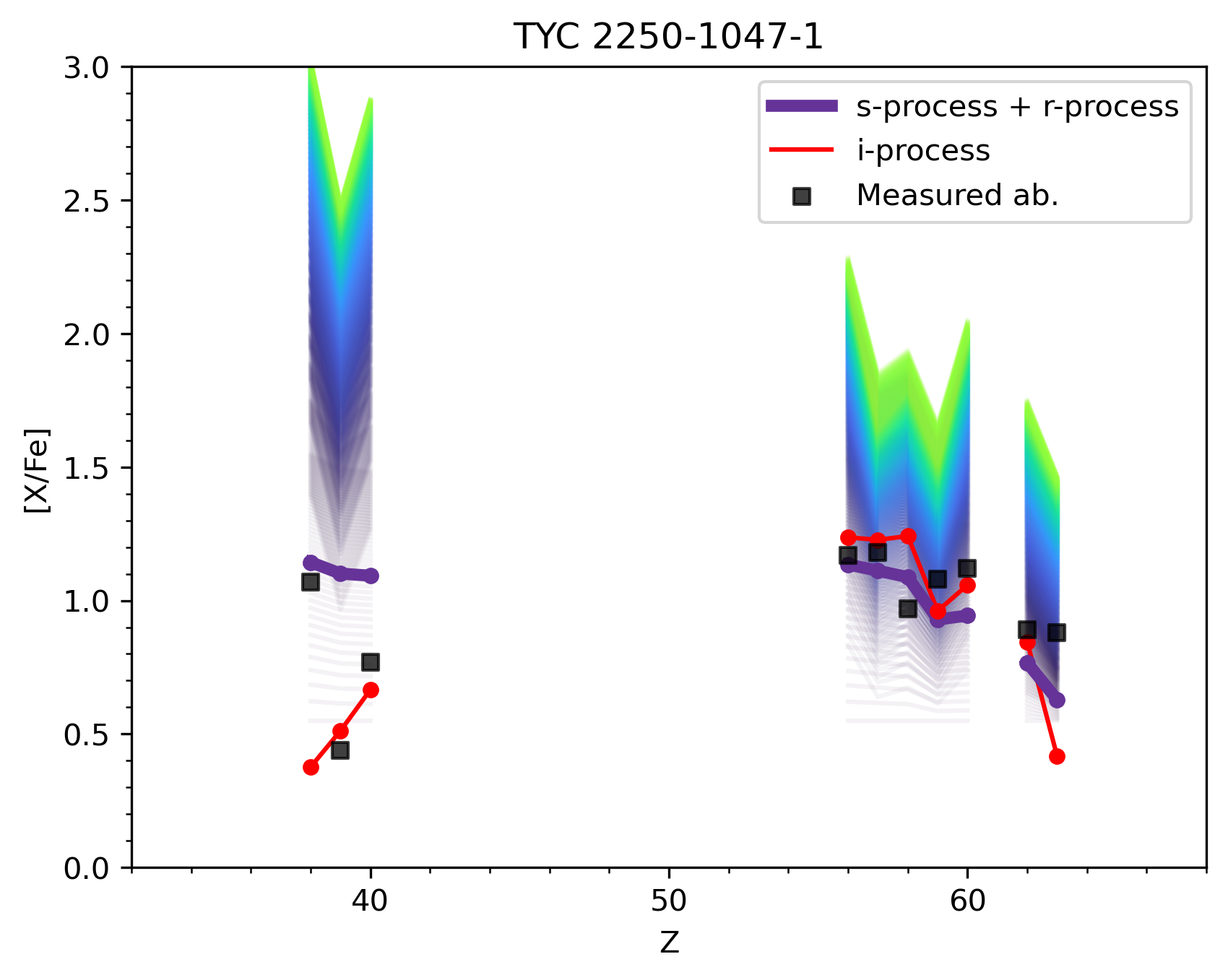}
\includegraphics[width=9cm]{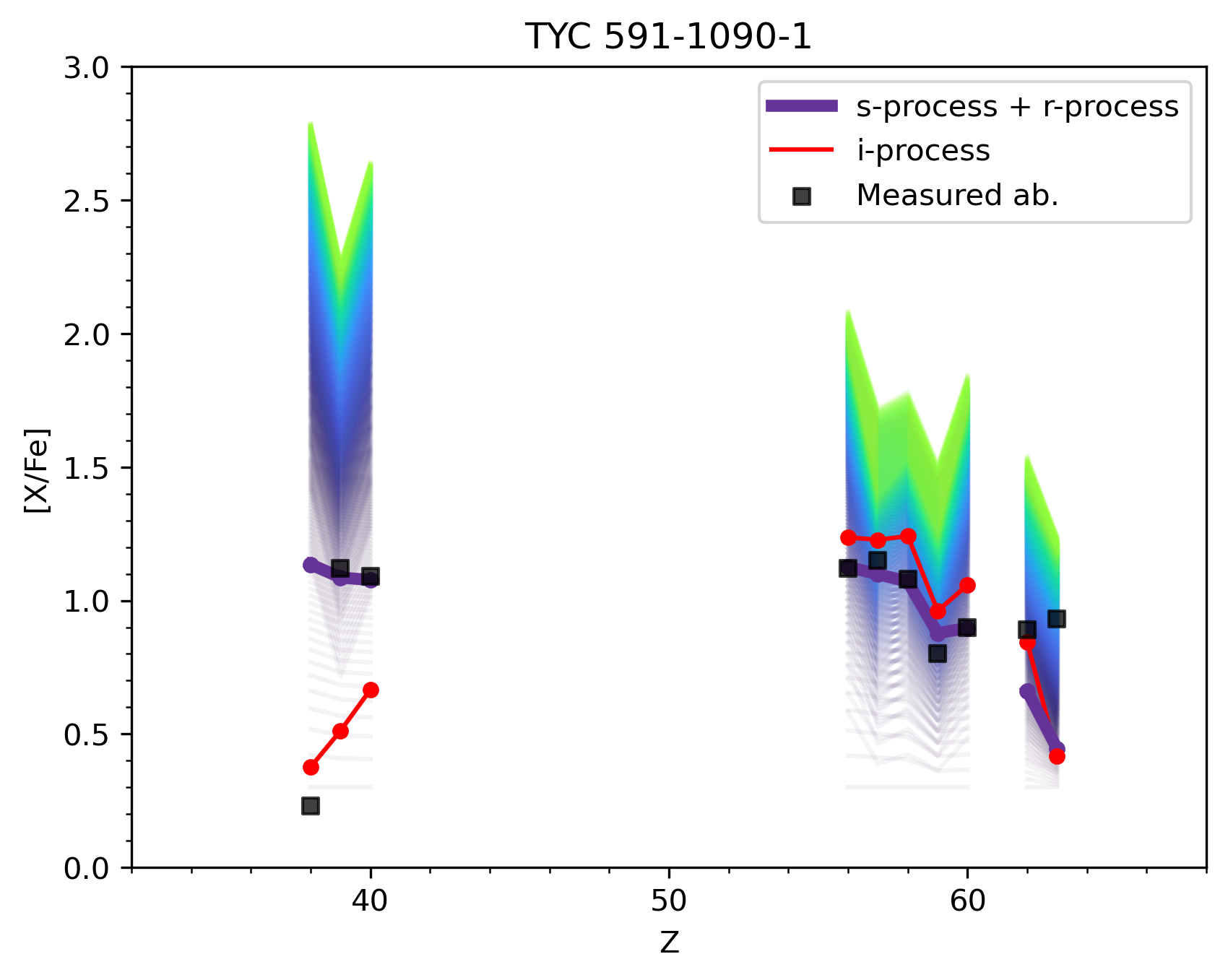}
\caption{Comparison of the measured abundance profiles (black squares) with a series of synthetic profiles made by a superposition of s- (as plotted in blue in Fig.~\ref{Fig:pattern}) and  r-process profiles, both varied independently and continuously from a solar-scaled profile with a very low s+r enrichment (purple faded lines) to large s+r enrichments (green faded lines). 
The synthetic "s+r" profile best matching the measured abundances is plotted as a thick violet line.  
For both stars it actually corresponds to a pure s-process contribution (hence any r-process contribution would deteriorate the agreement, whatever the s-process contribution). For reference, the i-process profile best matching the measured abundances is plotted in red as in Fig.\ref{Fig:pattern}. 
}

\label{Fig:sandr}
\end{figure}

To check that the two i-process star (TYC 2250-1047-1 and TYC 591-1090-1)
abundance patterns
are better fitted with an i-process than with a superposition of two independent pollutions, one by an s-process and another one by an r-process, we performed the following test.
A mixed r+s profile was added to a solar abundance profile scaled to the metallicity of the considered star.
The r+s profile was built
summing an r- and an s-contribution,
each one varied independently from zero value to values producing overabundances largely exceeding the measured abundances. 
We assume a solar-like pattern for the r-process
\citep[Table~B.4 of][]{karinkuzhi2021}
and an AGB pattern for the s-process at the corresponding stellar metallicity (blue line in Fig.~\ref{Fig:pattern}; see also
\citealt{Goriely18c}).
The agreement between the resulting abundance profile (plotted in faded colors in Fig.~\ref{Fig:sandr})
and the measured profile was quantified by computing a $\chi^2$ as in Eq.~\ref{eq:chi2}.
For the two stars, a pure s-process 
produces a better agreement with the measured abundances than a r+s mixture.
An r-process contribution is never favored because the slope of the r-process is quite steep (i.e., it is more rapidly decreasing with increasing $Z$ than the s-process), so matching the high europium (as well as samarium) abundances would 
inevitably lead to a large overestimate of the light-s (Sr, Y, Zr) abundances.
The resulting $\chi^2_{s}$ (between the s-profile and the measured abundances) is very similar to the $\chi^2_{i}$ 
(between the i-profile and the measured abundances), but the s-profile never explains the high europium abundance measured in TYC 2250-1047-1 and TYC 591-1090-1.
The present analysis indicates that these two enriched stars are better explained by an i-process than by a superposition of two independent s and r processes.

%******

%Hence the i-process better explains the abundances of  the two most enriched stars than a combination of s and r contributions. 
Some deviations between the measured and i-process abundances can still be noted, in particular for light s-elements, but also for Eu, which, in contrast to Sm, tends to be underestimated by the model. The absence of odd-even effect between measured Sm-Eu abundance is puzzling.

\section{Conclusions}
\label{Sect:conclusions}

Our re-analysis of fifteen stars classified as s-process-enriched by \cite{Norfolk2019} indicates that their abundances seem to be rather characterized by a hybrid sr pattern, rather than by a pure s-process pattern, as inferred from the distance indicator ($d_s$) that quantifies the deviation of the star's abundance profile from a pure r-process distribution.
For the two most enriched stars of our sample 
(that have [Eu/Fe]$\gtrsim0.75$ and that cannot be explained by Galactic chemical evolution models)
the comparison of the abundance profile with detailed calculations show that an i-process nucleosynthesis (resulting from the occurrence of a PIE) fits better the observed abundances than a pure s-process or than any combination of r- and s-patterns.
At low-enrichment levels, the primordial composition resulting from the galactic chemical evolution blurs the picture. 
This can be seen in Fig.~\ref{Fig:2}, where the barium stars enter the r+s region below [s/Fe]=0.5 dex.

It is possible that a non-negligible fraction of the stars previously identified as barium stars (because barium enhancement is relatively easy to identify) also present an r-element enrichment, but it is only recently that r-element abundances have been systematically investigated among barium stars.
A population of rs-enriched objects 
%\LS{with [Eu/Fe] \ga 0.5} 
(with [La/Eu] < 0.5)
seems to emerge, not only at low-metallicities where it was first noticed (among CEMP-rs stars), but also at close-to-solar metallicities, as also found in other samples of barium stars \citep{Karinkuzhi2018,denHartogh2022} and in Sakurai's object \citep{Herwig2011, vanHoof-2017}.
If the i-process is indeed active in objects like AGB stars of close-to-solar metallicity, important impact on nucleosynthesis and galactic chemical evolution can be expected. In particular, the i-process contribution to our solar system would need to be revisited.

\begin{acknowledgements}
 D.K. acknowledges the financial support from University of Calicut through seed money grant. SVE thanks the Fondation ULB for its support. 
 The {\it Mercator} telescope is operated thanks to grant number G.0C31.13 of the FWO under the “Big Science” initiative of the Flemish governement. Based on observations obtained with the HERMES spectrograph, supported by the Fund for Scientific Research of Flanders (FWO), the Research Council of K.U.Leuven, the Fonds National de la Recherche Scientifique (F.R.S.- FNRS), Belgium, the Royal Observatory of Belgium, the Observatoire de Genève, Switzerland and the Thüringer Landessternwarte Tautenburg, Germany.  LS and SG are senior research associates from F.R.S.- FNRS (Belgium). This work was supported by the F.R.S.- FNRS Grant No IISN 4.4502.19.

\end{acknowledgements}

\bibliographystyle{aa}
\bibliography{references}

\begin{appendix}

\section{Individual abundances}
Table~\ref{Tab:abundances} lists the elemental abundances.

%\renewcommand{\thetable}{A\arabic{table}}
%\renewcommand{\thetable}{\Alph{chapter}\arabic{table}}
%\setcounter{table}{0}
%{\footnotesize
\begin{table*}\small
\caption{Elemental abundances}
\label{Tab:abundances}
\begin{tabular}{lclllrllrrlrlrr}  
\hline
\\
\multicolumn{2}{c}{}& \multicolumn{4}{c}{BD $-07^\circ 402$} &&\multicolumn{3}{c}{BD $+44^\circ 575$}&& \multicolumn{3}{c}{HD 7863} \\
\cline{4-6}\cline{8-10}\cline{12-14}\\
 &    Z  &    log$_{\odot}{\epsilon}^a$ & log${\epsilon}$&$\sigma_{l}$ (N)& [X/Fe] $\pm~\sigma_{t}$ & & log${\epsilon}$&$\sigma_{l}$(N)& [X/Fe]~$\pm~ \sigma_{t}$ &&  log${\epsilon}$&$\sigma_{l}$(N)& [X/Fe]~$\pm~ \sigma_{t}$ \\
 &       &                         &                  &        &        \\

\hline
\ion{Pr}{II} & 59 & 0.72   & 0.70  & 0.10(3)& 0.09 $\pm$0.12 &&  0.70  & 0.10(1) & 0.43 $\pm$ 0.14  &&0.43 & 0.03(2) & $-$0.22 $\pm$ 0.10 &  \\
\ion{Nd}{II} & 60 & 1.42   & 1.49  & 0.09(8)& 0.18 $\pm$ 0.08  &&  1.42  & 0.06(7) & 0.45 $\pm$ 0.08   &&1.35 & 0.08(6) &  0.00 $\pm$ 0.08 &  \\ 
\ion{Sm}{II} & 62 & 0.96   & 1.08  & 0.10(5)& 0.23 $\pm$0.05 &&  0.96  & 0.06(4) & 0.45 $\pm$ 0.05   &&     &         &       &  \\
\ion{Eu}{II} & 63 &0.52    & 0.65  & 0.10(2)& 0.24 $\pm$ 0.09 &&  0.50  & 0.10(2) & 0.43 $\pm$ 0.09   &&0.50 & 0.10(4) &  0.05 $\pm$ 0.08 &  \\

\hline
\\
\multicolumn{2}{c}{}& \multicolumn{4}{c}{HIP 69788} && \multicolumn{3}{c}{TYC 22$-$155$-$1} && \multicolumn{3}{c}{TYC 2250$-$1047$-$1} \\
\cline{4-6}\cline{8-10}\cline{12-14}\\
 &    Z  &    log$_{\odot}{\epsilon}^a$ & log${\epsilon}$&$\sigma_{l}$(N)&  [X/Fe]~$\pm~ \sigma_{t}$ && log${\epsilon}$&$\sigma_{l}$(N)& [X/Fe]~$\pm~ \sigma_{t}$ && log${\epsilon}$&$\sigma_{l}$(N)& [X/Fe]~$\pm~ \sigma_{t}$ \\
 &       &                         &                  &        &        \\
\hline

\ion{Pr}{II} & 59 & 0.72 & 0.70 & 0.10(1) &0.02 $\pm$ 0.14   & & 0.70  & 0.10(3) &0.18 $\pm$ 0.12  &&  1.15 & 0.10(3) & 1.08 $\pm$ 0.12 &  \\
\ion{Nd}{II} & 60 & 1.42 & 1.20 & 0.10(3) &$-$0.18 $\pm$ 0.10 & & 1.55  & 0.06(7) &0.28 $\pm$ 0.08  &&  1.99 & 0.09(13)& 1.12 $\pm$ 0.08 &  \\
\ion{Sm}{II} & 62 & 0.96 & 1.00 & 0.10(3) &0.08 $\pm$ 0.09  & & 0.98  & 0.04(5) &0.22 $\pm$ 0.05  &&  1.30 & 0.10(4) & 0.89 $\pm$ 0.06 &  \\
\ion{Eu}{II} & 63 & 0.52 & 0.20 & 0.10(2) &$-$0.28 $\pm$ 0.09 & & 0.80  & 0.10(2) &0.48 $\pm$ 0.09  &&  0.85 & 0.08(3) & 0.88 $\pm$ 0.08 &  \\
\hline
\\
\multicolumn{2}{c}{}& \multicolumn{4}{c}{TYC 2913$-$1375$-$1} &&\multicolumn{3}{c}{TYC 2955$-$408$-$1}&& \multicolumn{3}{c}{TYC 3144$-$1906$-$1} \\
\cline{4-6}\cline{8-10}\cline{12-14}\\
 &    Z  &    log$_{\odot}{\epsilon}^a$ & log${\epsilon}$&$\sigma_{l}$ (N)& [X/Fe] $\pm~\sigma_{t}$ & & log${\epsilon}$&$\sigma_{l}$(N)& [X/Fe]~$\pm~ \sigma_{t}$ &&  log${\epsilon}$&$\sigma_{l}$(N)& [X/Fe]~$\pm~ \sigma_{t}$ \\
 &       &                         &                  &        &        \\
\hline
\ion{Pr}{II} & 59 & 0.72   & 0.27 & 0.05(3) & 0.16 $\pm$ 0.11 & & 0.95 & 0.04(3) & 0.62 $\pm$ 0.11   && 0.70 & 0.10(2) &0.11 $\pm$ 0.13   &   \\
 \ion{Nd}{II}& 60 & 1.42   & 0.96 & 0.08(9) & 0.15 $\pm$ 0.08 & & 1.69 & 0.13(7) & 0.66 $\pm$ 0.09  && 1.30 & 0.03(9) & 0.01 $\pm$ 0.08   &   \\
\ion{Sm}{II} & 62 & 0.96   & 0.68 & 0.12(6) & 0.33 $\pm$ 0.06 & & 1.27 & 0.05(7) & 0.70 $\pm$ 0.05   && 1.00 &0.20(6)  & 0.17 $\pm$ 0.09   &   \\
\ion{Eu}{II} & 63 & 0.52   & 0.30 & 0.10(2) & 0.39 $\pm$ 0.09 & & 0.65 & 0.05(2) & 0.52 $\pm$ 0.06   && 0.55 &0.07(3)  & 0.16 $\pm$ 0.07   &   \\
\hline
\\
\multicolumn{2}{c}{}& \multicolumn{4}{c}{TYC 3305$-$571$-$1} && \multicolumn{3}{c}{TYC 3423$-$6966$-$1} && \multicolumn{3}{c}{TYC 4684$-$2242$-$1} \\
\cline{4-6}\cline{8-10}\cline{12-14}\\
 &    Z  &    log$_{\odot}{\epsilon}^a$ & log${\epsilon}$&$\sigma_{l}$(N)&  [X/Fe]~$\pm~ \sigma_{t}$ && log${\epsilon}$&$\sigma_{l}$(N)& [X/Fe]~$\pm~ \sigma_{t}$ && log${\epsilon}$&$\sigma_{l}$(N)& [X/Fe]~$\pm~ \sigma_{t}$ \\
 &       &                         &                  &        &        \\
\hline

\ion{Pr}{II} & 59 & 0.72 & 0.85 & 0.15(2)  & 0.18 $\pm$ 0.15   && 0.85  & 0.10(1) & 0.11 $\pm$ 0.14 && 0.70 & 0.10(2) & 0.03 $\pm$ 0.12  &   \\
\ion{Nd}{II} & 60 & 1.42 & 1.67 & 0.05(3)  & 0.30 $\pm$ 0.09  && 1.65  & 0.10(6) & 0.21 $\pm$ 0.09 && 1.50 & 0.10(6) & 0.13 $\pm$ 0.09  &   \\
\ion{Sm}{II} & 62 & 0.96 & 1.25 & 0.05(3)  & 0.34 $\pm$ 0.05   && 1.00  & 0.05(3) & 0.02 $\pm$ 0.05 && 0.98 & 0.04(5) & 0.07 $\pm$ 0.05  &   \\
\ion{Eu}{II} & 63 & 0.52 & 0.80 & 0.10(2)  & 0.33 $\pm$ 0.09   && 0.80  & 0.10(1) & 0.26 $\pm$ 0.12&& 0.80 & 0.10(2) & 0.33 $\pm$ 0.09 &   \\
\hline
\\
\multicolumn{2}{c}{}& \multicolumn{4}{c}{TYC 4837$-$925$-$1} &&\multicolumn{3}{c}{TYC 591$-$1090$-$1}&& \multicolumn{3}{c}{TYC 752$-$1944$-$1} \\
\cline{4-6}\cline{8-10}\cline{12-14}\\
 &    Z  &    log$_{\odot}{\epsilon}^a$ & log${\epsilon}$&$\sigma_{l}$ (N)& [X/Fe] $\pm~\sigma_{t}$ & & log${\epsilon}$&$\sigma_{l}$(N)& [X/Fe]~$\pm~ \sigma_{t}$ &&  log${\epsilon}$&$\sigma_{l}$(N)& [X/Fe]~$\pm~ \sigma_{t}$ \\
 &       &                         &                  &        &        \\
\hline
\ion{Pr}{II} & 59 & 0.72 & 0.52 & 0.06(3) & 0.07 $\pm$ 0.09  && 1.30 & 0.10(3) & 0.80 $\pm$ 0.12  &&  1.17 & 0.09(3)  & 0.53 $\pm$ 0.12    & \\
\ion{Nd}{II} & 60 & 1.42 & 1.33 & 0.12(7) & 0.18 $\pm$ 0.09  && 2.02 & 0.14(8) & 0.90 $\pm$ 0.09  &&  1.98 & 0.05(9)  & 0.64 $\pm$ 0.08    & \\
\ion{Sm}{II} & 62 & 0.96 & 0.80 & 0.12(4) & 0.11 $\pm$ 0.07  && 1.55 & 0.09(4) & 0.89 $\pm$ 0.06 &&  1.34 & 0.10(8)  & 0.46 $\pm$ 0.05    & \\
\ion{Eu}{II} & 63 & 0.52 & 0.50 & 0.10(2) & 0.25 $\pm$ 0.09  && 1.15:& 0.10(1) & 0.93 $\pm$ 0.12  &&  0.80 & 0.10(2)  & 0.36 $\pm$ 0.09    & \\

\hline
\end{tabular}

$^{a}$ Asplund et al. (2009) \\
$:$ Uncertain abundances due to noisy/blended region\\

\end{table*}

\section{Abundance pattern in program stars}
\begin{figure}
\includegraphics[width=9cm]{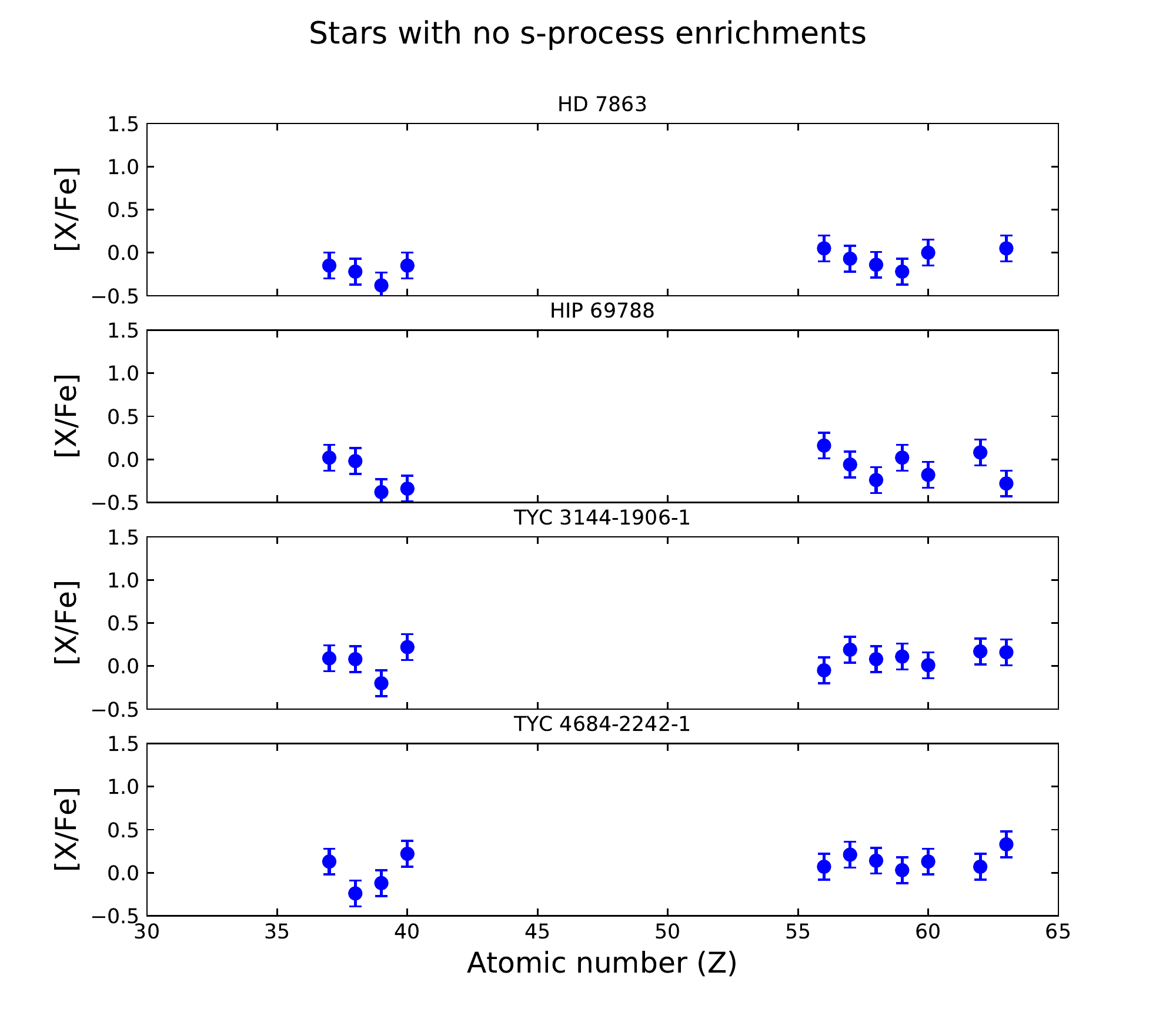}
\includegraphics[width=9cm]{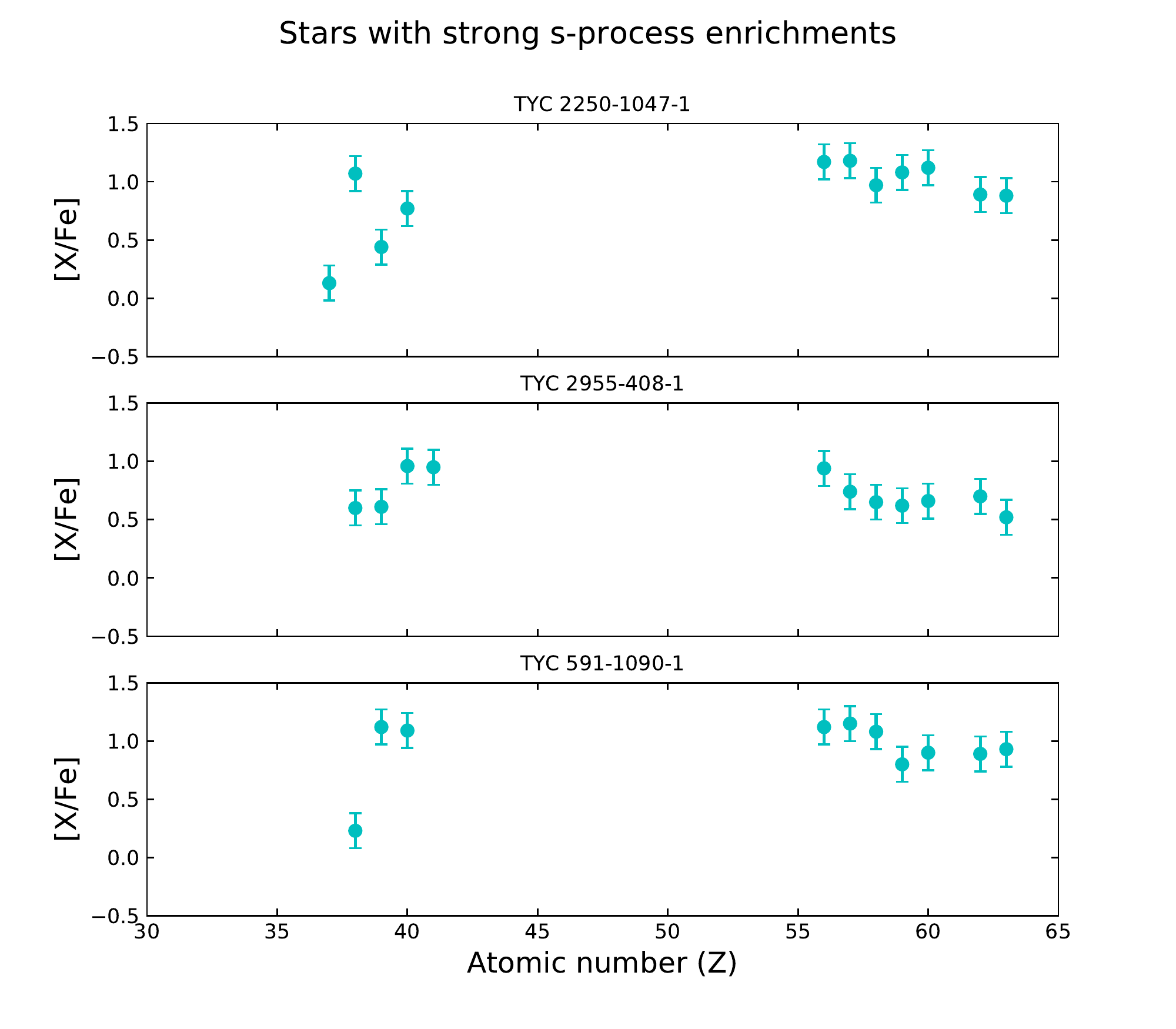}
\caption{Abundance profiles of the program stars color - coded according to the adopted classification criteria. The upper panel shows the abundance pattern in stars with no s-process enhancements. The lower panel shows the the abundance pattern in three stars with strong s-process enrichments. }
\label{Fig:pattern2}
\end{figure}

\begin{figure}
\includegraphics[width=9cm]{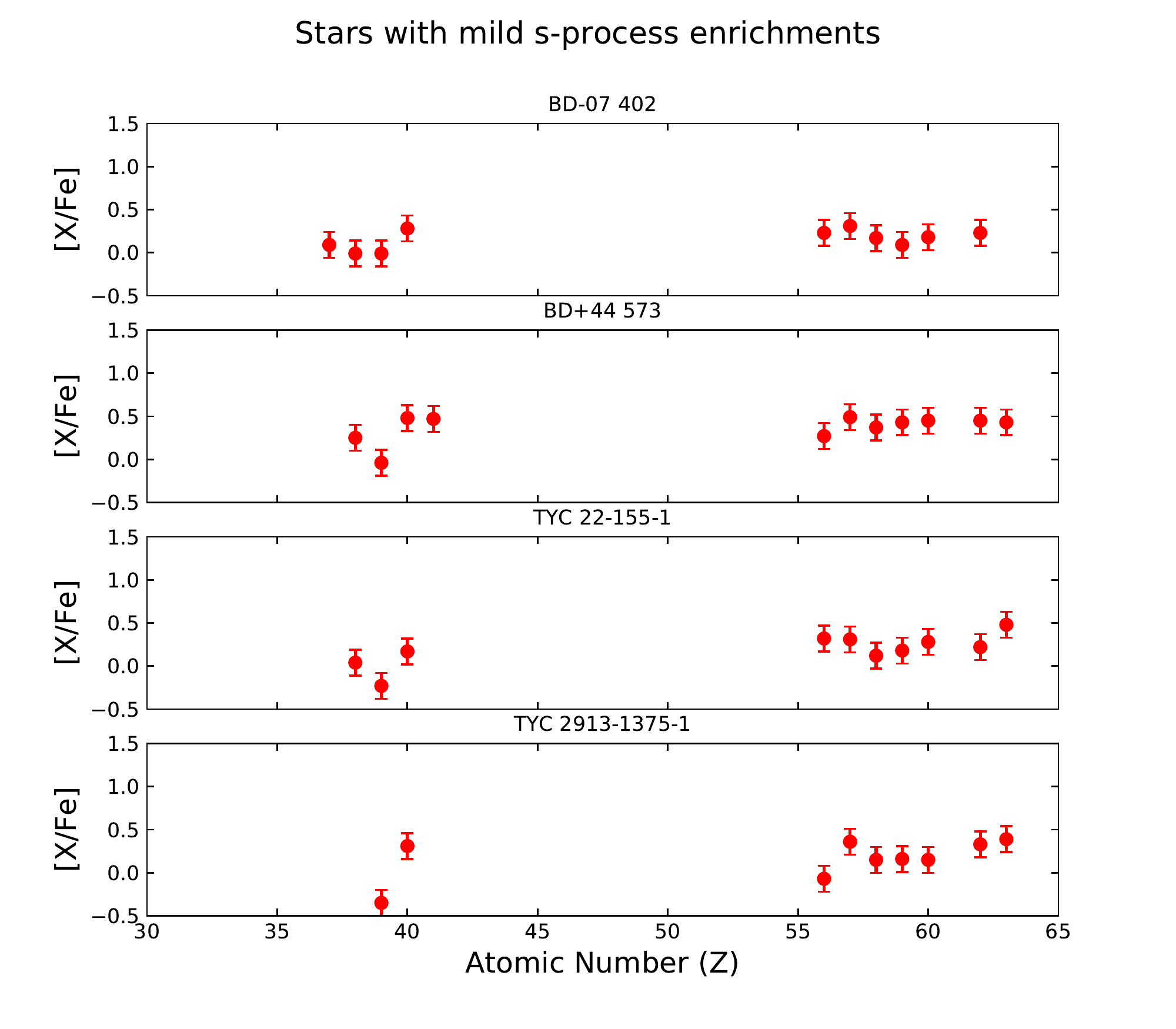}
\includegraphics[width=9cm]{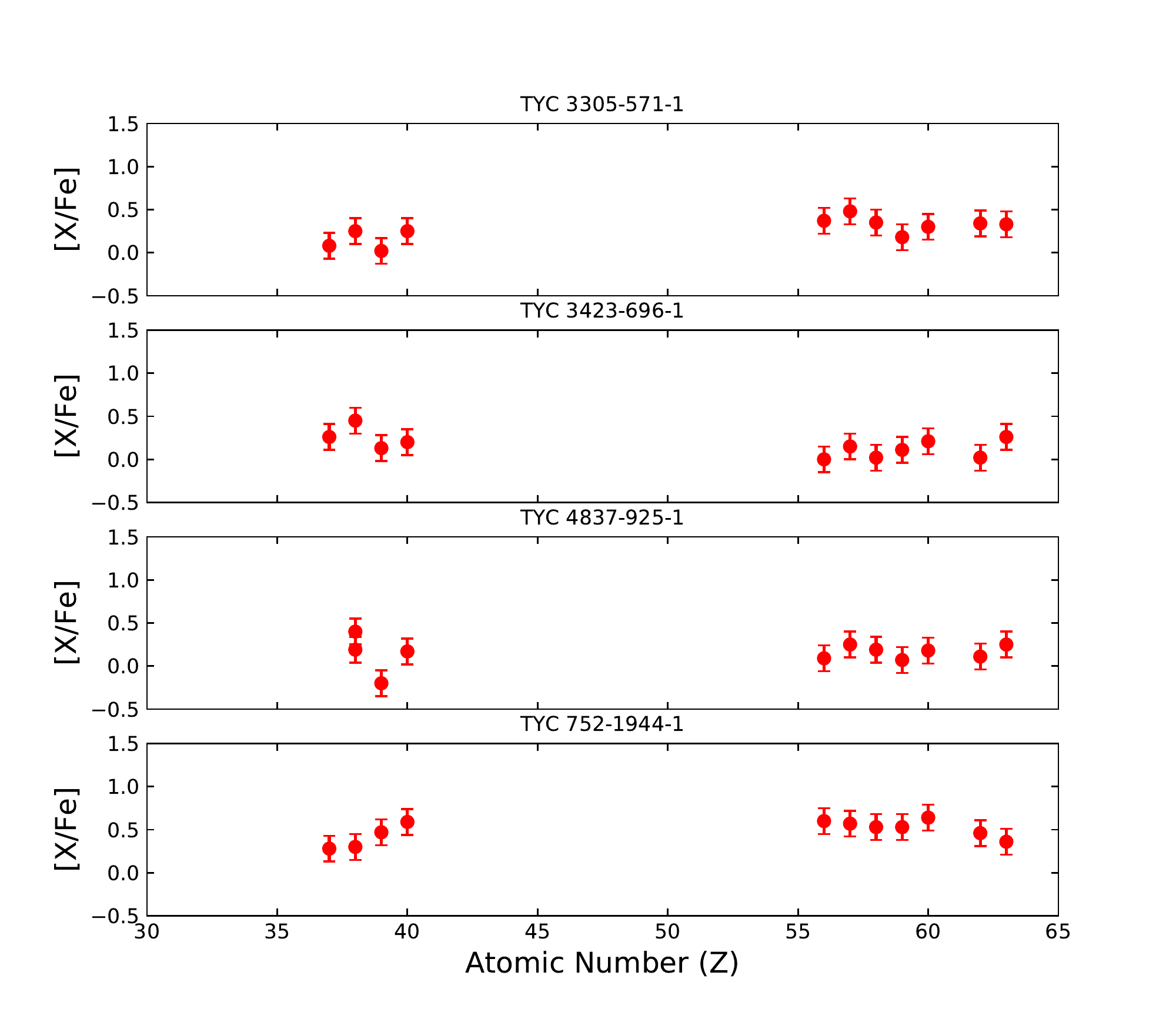}

\caption{Same as \ref{Fig:pattern2} but for the the eight mild s-process enriched stars. }
\label{Fig:pattern3}
\end{figure}

Fig.~\ref{Fig:pattern2} and ~\ref{Fig:pattern3} show the abundance patterns for stars showing no, mild or strong s-process enrichments. Figure~\ref{Fig:1} 
displays the ([Ba/Fe], [Eu/Fe]) diagram.


\end{appendix}
\end{document}